# Lower Bound On the Computational Complexity of Discounted Markov Decision Problems

Yichen Chen,* Mengdi Wang†

May 20, 2017

## Abstract

We study the computational complexity of the infinite-horizon discounted-reward Markov Decision Problem (MDP) with a finite state space $\mathcal{S}$ and a finite action space $\mathcal{A}$. We show that any randomized algorithm needs a running time at least $\Omega(|\mathcal{S}|^2|\mathcal{A}|)$ to compute an $\epsilon$-optimal policy with high probability. We consider two variants of the MDP where the input is given in specific data structures, including arrays of cumulative probabilities and binary trees of transition probabilities. For these cases, we show that the complexity lower bound reduces to $\Omega\left(\frac{|\mathcal{S}||\mathcal{A}|}{\epsilon}\right)$. These results reveal a surprising observation that the computational complexity of the MDP depends on the data structure of input.

## 1 Introduction

The Markov Decision Problem (MDP) arises from stochastic control processes in which a planner aims to make a sequence of decisions as the state of the process evolves. It provides a basic mathematical framework for dynamic programming and reinforcement learning. It finds wide applications in engineering systems, operations research, artificial intelligence and computer games.

Consider the infinite-horizon discounted-reward MDP with finitely many states and finitely many actions. An instance of such MDP can be described by a tuple $\mathcal{M} = (\mathcal{S}, \mathcal{A}, P, r, \gamma)$ where $\mathcal{S}$ is a finite state space, $\mathcal{A}$ is a finite action space , $\gamma \in (0, 1)$ is the discount factor , $r : \mathcal{S} \times \mathcal{A} \mapsto [0, 1]$ is the reward function and $P = (P_a)_{a \in \mathcal{A}}$ is the collection of matrices of transition probabilities. The size of a MDP model is $\mathcal{O}(|\mathcal{S}|^2|\mathcal{A}|)$.

The MDP tuple $\mathcal{M}$ specifies a controlled random walk on the state space $\mathcal{S}$. When action $a \in \mathcal{A}$ is selected at state $i \in \mathcal{S}$, the system transitions to a state $j \in \mathcal{S}$ with probability $P_a(i, j)$ and receives the reward $r_{ia}$. A (stationary) deterministic policy $\pi : \mathcal{S} \mapsto \mathcal{A}$ is a mapping from states to actions. A randomized policy can be viewed as a collection of probability distributions for choosing actions at each state. The value vector $v^\pi \in \mathcal{R}^{|\mathcal{S}|}$ under a policy $\pi$ is defined as

$$v^\pi(i) = \mathbf{E}^\pi \left[ \sum_{k=0}^{\infty} \gamma^k r_{i_k \pi(i_k)} \mid i_0 = i \right], \qquad (1)$$

where $(i_0, i_1, \dots)$ is the path of state transitions generated by the Markov chain under policy $\pi$ and the expectation is taken over the path. It is known [Bel57] that there exists an optimal deterministic policy that maximizes the value vector for all states simultaneously, yielding the optimal value vector $v^* = \max_\pi v^\pi$. Our objective is to find a (stationary and possibly randomized) policy that performs nearly well as the optimal policy. We say that a policy $\pi$ is $\epsilon$-optimal if $\|v^* - v^\pi\|_\infty \leq \epsilon$.

In this paper, we study the computational complexity of finding an $\epsilon$-optimal policy of the MDP. We focus on the running-time complexity, which is the sum of the number of arithmetic operations (including addition, subtraction, multiplication and division) and the number of queries to input data (each query asks for one specified entry of the input). Although the MDP is a classical problem that has been studied for years, there has been few results on its computational complexity lower bound. We have not been able to find comparable existing work. To our best knowledge, this is the first work on lower bounds of the computational complexity for finite-state finite-action MDP.

---

*Yichen Chen is with Department of Computer Science, Princeton University, Princeton 08544, USA.

†Mengdi Wang is with Department of Operations Research and Financial Engineering, Princeton University, Princeton 08544, USA.

Our main results are summarized as follows:

1. We show that for the standard discounted MDP, the running time needed to compute an $\epsilon$-optimal policy with probability at least 9/10 is
$$\Omega(|\mathcal{S}|^2|\mathcal{A}|).$$
as long as $\epsilon < \frac{\gamma}{4(1-\gamma)}$. This result asserts that the running-time complexity of MDP is at least *linear* with respect to the input size $\mathcal{O}(|\mathcal{S}|^2|\mathcal{A}|)$.

2. We also consider two variants of the discounted MDP where the input is given in specific data structures: arrays of cumulative sums or binary trees. In these two cases, we show that the running time needed to compute an $\epsilon$-optimal policy with probability at least 9/10 is
$$\Omega\left(\frac{\gamma}{1-\gamma} \cdot \frac{|\mathcal{S}||\mathcal{A}|}{\epsilon}\right).$$
This result suggests that the complexity of MDP reduces when the input data is given in convenient data structures. It leaves open the possibility for obtaining approximate solutions in *sublinear* running time.

Similar phenomenon has been observed in a recent paper that studies the running-time complexity upper bounds of discounted MDP [Wan17]. In [Wan17], an upperbound $\tilde{\mathcal{O}}\left(|\mathcal{S}|^3|\mathcal{A}|\right)$ was established for the standard MDP using a randomized primal-dual algorithm. The upper bound was shown to reduce to $\tilde{\mathcal{O}}\left(\frac{|\mathcal{S}||\mathcal{A}|}{\epsilon^2}\right)$ under some additional assumption on the ergodicity when the input is given in convenient data structures (i.e., sorted arrays, cumulative probabilities, and binary trees). The main advantage of these data structures is that they enable immediate simulation of the Markov process in the algorithm, requiring zero preprocessing time.

Let us compare our lower bound results with the upper bounds mentioned above. Although there remains a gap between the upper and lower bounds, there seems to be a sharp contrast between the case with standard input and the case where the input data is given in convenient data structures. This contrast leads us to a counter intuitive conjecture:

*The computational complexity for solving the MDP depends on the data structure of the input.*

More importantly, the nice data structures that lead to a lower computational complexity happen to be those that enable immediate sampling of the Markov state transitions. This observation draws an intriguing connection between the running-time complexity for solving the MDP and the sample complexity for estimating the optimal policy. Yet this connection is not fully understood. We believe that this work on complexity lower bounds for the MDP will provide critical insights into the complexity of sequential decision-making problems and reinforcement learning. We hope these results will motivate sharper analysis and better algorithms.

## 2 Related Works

Complexity analysis for the MDP has a long history started by Bellman [Bel57]. The MDP was known to be solvable in polynomial time by dynamic programming [PT87]. Yet the complexity's dependence on the size of the state space and the action space is not clear.

There are three major approaches for solving the MDP: the value iteration method, the policy iteration method, and the linear programming method. Most of the existing works focus on establishing the upper bound of the computational complexity for these methods; see e.g., [How60, Tse90, LDK95, MS99, Ye05,

Ye11, Sch13, FH14]. The best known complexity results for these methods are superlinear with respect to the input size $\mathcal{O}(|\mathcal{S}|^2|\mathcal{A}|)$. Recently, [Wan17] developed a randomized primal-dual method that achieves a sublinear running time $\mathcal{O}\left(\frac{|\mathcal{S}||\mathcal{A}|}{\epsilon^2}\right)$ under additional assumptions on the ergodicity and the format of the MDP input.

In contrast to the large volumes of upper bound results, there are fewer results on the complexity lower bound. Papadimitriou and Tsitsiklis [PT87] showed that the MDP is complete for $P$ and its partial information variant is $PSPACE$-complete. Chow and Tsitsiklis [CT89] established a complexity lower bound of $\Omega(\frac{1}{\epsilon^{2n+m}})$ for continuous-state Markov decision process under some smoothness conditions, where $n$, $m$ are the dimensions of the continuous state and action spaces respectively. Blondel and Tsitsiklis [BT00] surveyed basic results on the complexity of the MDP and more general stochastic control problems. Friedman, Hansen and Zwich [FHZ11] established subexponential lower bound for the simplex method applied to MDP using randomized pivoting rules. Hansen and Zwich [HZ10] showed that Howard's policy iteration algorithms require at least $\Omega(|\mathcal{S}|^2)$ iterations for the deterministic MDP with average cost. Yet most of these results apply to specific algorithms. We are interested in developing a lower bound that works for arbitrary algorithms. To our best knowledge, such a result is missing in the literature.

Despite the lack of existing results on the computational complexity, a related notion called sample complexity has been widely studied in the setting of reinforcement learning; e.g., [EDMM02, MT04, SLW+06, SLL09, AMK12, LH12, DB15]. Sample complexity is the number of state transitions one needs to observe in order to estimate the optimal policy. In existing works on sample complexity, although the settings and assumptions vary from one to another, the sample complexity results are typically $\Omega(\frac{|\mathcal{S}||\mathcal{A}|}{\epsilon^2})$. Apparently the notion of sample complexity is fundamentally different from the notion of running-time complexity. Thus the corresponding results cannot be directly compared.

Our proof of the lower bound for computational complexity is inspired by works in theoretical computer science. For example, Clarkson, Hazan and Woodruff [CHW12] shows that in order to achieve $\epsilon$-optimality for linear classification problem, the algorithm needs to query at least $\Omega(\epsilon^{-2}(m+n))$ entries of the $m$ by $n$ input matrix. Our analysis involves constructing a tree program, which shares a similar spirit as [BFK+81, BFMadH+87, Yao94, Raz16]. These works study the time-space tradeoff for computation problems such as sorting and parity learning, while our work focuses on the time complexity for the MDP and assumes adequate space. Our analysis is also related to the relational adversary method proposed in [Amb00, Aar06].

## 3 Main Theorems

In this section, we establish the lower bound of computational complexity for any algorithm that takes the specification of an instance of MDP as input and outputs a policy. We will show that the format and data structure of the input is of vital importance. We consider three distinct cases: (1) the standard case where the transition probabilities $P$ are given as arrays; (2) the case where the cumulative probabilities are given instead of transition probabilities; (3) the case where the transition probabilities are given in the format of binary trees. We will show that the first case is more difficult than the latter two cases. The proofs are deferred to Sections 4, 5 and Appendix.

First we consider the standard case and define `Standard` MDP as follows.

**Definition 1** (Standard MDP). *The input of `Standard` MDP includes:*

- *Transition probability matrices $P_a$ of dimension $|\mathcal{S}| \times |\mathcal{S}|$, for all $a \in \mathcal{A}$.*
- *A reward vector $r$ of dimension $|\mathcal{S}| \times |\mathcal{A}|$ and $\epsilon > 0$.*

This is the standard format of the input that is widely used for solving the MDP [Ber95, BT95, Put14, Ber13]. For `Standard` MDP, we establish the following lower bound on the running time of any (possibly randomized) algorithm.

**Theorem 1.** *For $\epsilon \leq \frac{\gamma}{4(1-\gamma)}$, any randomized algorithm for `Standard` MDP needs a running time at least $\Omega(|\mathcal{S}|^2|\mathcal{A}|)$ to compute an $\epsilon$-optimal policy with probability at least $9/10$.*

In Theorem 1, the condition $\epsilon \leq \frac{\gamma}{4(1-\gamma)} \approx \frac{1}{4(1-\gamma)}$ essentially indicates a *constant multiplicative approximation ratio* that is close to $1/4$, because $v^\pi$ is typically on the order of $\frac{1}{1-\gamma}$. It says that, in order to get a reasonable approximate policy, the running time is at least linear with respect to the input size $\mathcal{O}(|\mathcal{S}|^2|\mathcal{A}|)$. This lower bound might seems quite intuitive, however, the formal proof requires the use of Yao's minimax principal and careful analysis of the tree program associated with any algorithm. The result of Theorem 1 excludes the possibility of the existence of any sublinear running-time algorithm for standard MDP.

Second we consider the case where transition probabilities of MDP are implicitly given in the format of cumulative probabilities (assuming that the states are ordered in some way). We define `CDP MDP` as follows.

**Definition 2** (MDP Specified Using Cumulative Probabilities). *The input of `CDP MDP` includes:*

- *Matrices of cumulative probabilities $C_a$ of dimension $|\mathcal{S}| \times |\mathcal{S}|$, for all $a \in \mathcal{A}$, where $C_a(i,j) = \sum_{k=1}^{j} P_a(i,k)$ for all $i \in \mathcal{S}, a \in \mathcal{A}$ and $j \in \mathcal{S}$.*

- *A reward vector $r$ of dimension $|\mathcal{S}| \times |\mathcal{A}|$ and $\epsilon > 0$.*

The following theorems concerns the lower bound for solving `CDP MDP`.

**Theorem 2.** *For $\epsilon \geq 3/|\mathcal{S}|$, any algorithm for `CDP MDP` needs at least $\Omega\left(\frac{|\mathcal{S}||\mathcal{A}|}{\epsilon}\right)$ running time to compute an $\frac{\epsilon\gamma}{4(1-\gamma)}$-optimal policy with probability at least $9/10$.*

According to Theorem 2, the running time needed to compute an $\epsilon$-optimal policy with probability at least $9/10$ is $\Omega\left(\frac{|\mathcal{S}||\mathcal{A}|\gamma}{\epsilon(1-\gamma)}\right)$ as long as $\epsilon \geq \frac{3\gamma}{4|\mathcal{S}|(1-\gamma)}$. Note that this is a weaker lower bound than that of Theorem 1. It does not exclude the existence of sublinear-time approximation algorithms.

Third, we consider the case when the transition probabilities are given explicitly but are in the format of binary trees rather than arrays. We define the computation problem as follows.

**Definition 3** (MDP Specified Using Binary Trees). *The input of `Binary Tree MDP` includes:*

- *Transition probability distributions $P_a(i,\cdot)$ that are encoded in binary trees. There are $|\mathcal{S}||\mathcal{A}|$ trees, one for each state-action pair $(i,a) \in \mathcal{S} \times \mathcal{A}$. Each binary tree has $\log|\mathcal{A}|$ layers and $|\mathcal{S}|$ leaves that store the values of $P_a(i,j), j \in \mathcal{S}$. Each inner node of the tree stores the sum of its two children.*

- *A reward vector $r$ of dimension $|\mathcal{S}| \times |\mathcal{A}|$ and $\epsilon > 0$.*

We have the following computational complexity lower bound for solving `Binary Tree MDP`.

**Theorem 3.** *Any algorithm for `Binary Tree MDP` needs at least $\Omega\left(\frac{|\mathcal{S}||\mathcal{A}|}{\epsilon}\right)$ running time to produce an $\frac{\epsilon\gamma}{4(1-\gamma)}$-optimal policy with probability at least $9/10$ for $\epsilon$ larger than $3/|\mathcal{S}|$.*

The results of Theorems 2, 3 are similar. They suggest that the MDP becomes easier to solve when the input is given in specific formats. They provide a hint that one might be able to develop faster algorithms to exploit the structures of the input. One may wonder what is common to `CDP MDP` and `Binary Tree MDP`? The answer is quite interesting: both the input of `CDP MDP` and `Binary Tree MDP` allow immediate simulation of the Markov decision process without preprocessing [WE80].

In contrast, given the input of `Standard MDP`, one needs $\mathcal{O}(|\mathcal{S}|^2|\mathcal{A}|)$ running time to preprocess the arrays of transition probabilities in order to sample state-to-state transitions. This observation is consistent with the results of [Wan17] in which the upper bound reduces if the MDP input is given in a nice format that allows immediate sampling of state transitions. We conjecture that, once the Markov chain can be simulated, the MDP becomes easier to solve. This further implies a close connection between the running-time complexity of the MDP and the sample complexity of the associated reinforcement learning problem. This conjecture awaits further research.

# 4 Family of Hard Instances of MDP

In this section, we take a substantial step towards establishing the lower bound of computational complexity for the MDP. Let $\epsilon > 0$ be an arbitrary value. We will construct two disjoint sets of MDP instances that are close to each other. For any algorithm, the failure to distinguish one set from the other would result in a computation error at least $\epsilon$.

## 4.1 Hard Instances of `Standard MDP`

|  |  | $\mathcal{S}_\text{B}$ |  |  |  |  | $\mathcal{S}_\text{G}$ |  |  |  |  | $s_\text{N}$ |
|---|---|---|---|---|---|---|---|---|---|---|---|---|
| $M_1$ | $\mathcal{S}_\text{U} \times \mathcal{A}_U$ | 0 | 1 | $\cdots$ | 0 | 0 |  | 0 |  |  |  | 0 |
|  |  | 0 | 0 | $\cdots$ | 1 | 0 |  |  |  |  |  |  |
|  |  | $\vdots$ | $\vdots$ | $\ddots$ | $\vdots$ | $\vdots$ |  |  |  |  |  |  |
|  |  | 1 | 0 | $\cdots$ | 0 | 0 |  |  |  |  |  |  |
|  |  | 0 | 0 | $\cdots$ | 0 | 1 |  |  |  |  |  |  |
|  | $\mathcal{S}_\text{U} \times a_N$ | 0 |  |  |  |  |  | 0 |  |  |  | 1 |
| $M_2$ | $\mathcal{S}_\text{U} \times \mathcal{A}_U$ | 0 | 1 | $\cdots$ | 0 | 0 | 0 | 0 | $\cdots$ | 0 | 0 | 0 |
|  |  | 0 | 0 | $\cdots$ | 0 | 0 | 0 | 0 | $\cdots$ | 1 | 0 |  |
|  |  | $\vdots$ | $\vdots$ | $\ddots$ | $\vdots$ | $\vdots$ | $\vdots$ | $\vdots$ | $\ddots$ | $\vdots$ | $\vdots$ |  |
|  |  | 1 | 0 | $\cdots$ | 0 | 0 | 0 | 0 | $\cdots$ | 0 | 0 |  |
|  |  | 0 | 0 | $\cdots$ | 0 | 1 | 0 | 0 | $\cdots$ | 0 | 0 |  |
|  | $\mathcal{S}_\text{U} \times a_N$ | 0 |  |  |  |  |  | 0 |  |  |  | 1 |

Figure 1: Input of `Standard MDP`: Arrays of Transition Probabilities for $M_1 \in \mathcal{M}_1$ and $M_2 \in \mathcal{M}_2$.

Let the state space $\mathcal{S}$ consist of four parts, $\mathcal{S} = \mathcal{S}_\text{U} \cup \mathcal{S}_\text{G} \cap \mathcal{S}_\text{B} \cup \{s_\text{N}\}$, where $|\mathcal{S}_\text{U}| = |\mathcal{S}_\text{G}| = |\mathcal{S}_\text{B}| = \frac{|\mathcal{S}|-1}{3}$ and $s_\text{N}$ is a single state. Let the action space be $\mathcal{A} = \mathcal{A}_U \cup \{a_N\}$ where $|\mathcal{A}_U| = |\mathcal{A}| - 1$ and $a_N$ is a single action. We construct two sets of MDP instances $\mathcal{M}_1$ and $\mathcal{M}_2$ that are hard to distinguish, which are given below:

- Let $\mathcal{M}_1$ be the set of instances satisfying the following. If we select any action $a \in \mathcal{A}$ in state $s \in \mathcal{S}_\text{G}, \mathcal{S}_\text{B}$ or $s_\text{N}$, the state $s$ transitions to itself with a reward $1, 0$ or $1/2$. Intuitively, $\mathcal{S}_\text{G}, \mathcal{S}_\text{B}$ and $s_\text{N}$ are the good, bad and neutral states with high, low and median rewards. Given $a \in \mathcal{A}_U$ and $s \in \mathcal{S}_\text{U}$, the system transitions to some $s' \in \mathcal{S}_\text{B}$ with probability 1 and reward 0. Given $a_N$ and $s \in \mathcal{S}_\text{U}$, the state transitions to $s_\text{N}$ with reward 0. For $M_1 \in \mathcal{M}_1$, we have $v^*_{M_1}(s) = \frac{\gamma}{2(1-\gamma)}$ for all $s \in \mathcal{S}_\text{U}$. The cardinality of $\mathcal{M}_1$ is $|\mathcal{S}_\text{B}|^{|\mathcal{S}_\text{U} \times \mathcal{A}_U|}$.

- Let $\mathcal{M}_2$ be the set of instances that differ from those in $\mathcal{M}_1$ at one state-action pair, which we denote by $(\bar{s}, \bar{a}) \in \mathcal{S}_\text{U} \times \mathcal{A}_U$. Given $(\bar{s}, \bar{a})$, the system transitions to some good state $\bar{s}' \in \mathcal{S}_\text{G}$ with probability 1 and reward 0. For $M_2 \in \mathcal{M}_2$, we have $v^*_{M_2}(\bar{s}) = \frac{\gamma}{1-\gamma}$ and $v^*_{M_2}(s) = \frac{\gamma}{2(1-\gamma)}$ for all $s \in \mathcal{S}_\text{U}/\{\bar{s}\}$. The cardinality of $\mathcal{M}_2$ is $|\mathcal{S}_\text{U} \times \mathcal{A}_U| \times |\mathcal{S}_\text{B}|^{|\mathcal{S}_\text{U} \times \mathcal{A}_U|}$.

Figure 1 shows a pair of $M_1 \in \mathcal{M}_1$ and $M_2 \in \mathcal{M}_2$ which differ in two entries. We see that $\|v^*_{M_1} - v^*_{M_2}\|_\infty = \frac{\gamma}{2(1-\gamma)}$. Suppose that an algorithm can output a $\frac{\gamma}{4(1-\gamma)}$-optimal policy with high probability for all instances of the MDP. The algorithm must be able to differentiate $\mathcal{M}_1$ from $\mathcal{M}_2$ with high probability. Informally speaking, the algorithm needs to search for two discrepancies in an array of size $|\mathcal{S}_\text{U} \times \mathcal{A}_U \times \mathcal{S}| = \Omega(|\mathcal{S}|^2|\mathcal{A}|)$. Intuitively, it needs to query a significant portion of the entire array to succeed with high probability, which requires a running time of $\Omega(|\mathcal{S}|^2|\mathcal{A}|)$.

## 4.2 Hard Instances of `CDP MDP` and `Binary Tree MDP`

Let $\epsilon > 3/|\mathcal{S}|$ be arbitrary such that $1/\epsilon$ is an integer and let $k = (|\mathcal{S}|-1)/3$. We arrange the good states and the bad states in an alternating order, i.e., $s_{B,1}, s_{G,1}, \ldots, s_{B,k}, s_{G,k}$. Let us construct two sets of instances

$\mathcal{M}_3$ and $\mathcal{M}_4$ that are hard to distinguish:

- Let $\mathcal{M}_3$ consist of a single instance $M_3$ satisfying the follows. If we select action $a \in \mathcal{A}_U$ in state $s \in \mathcal{S}_U$, the system transitions to $s_{B,1}, \ldots, s_{B,1/\epsilon}$ each with probability $\epsilon$ and reward 0. From any $s \in \mathcal{S}_G, \mathcal{S}_B, s_N$, the system transitions to itself with a reward of $1, 0, \epsilon/2$ respectively, regardless of the action. If we select action $a_N$ at state $s \in \mathcal{S}_U$, it transitions to $s_N$ with probability 1 and reward 0. We have $v_{M_3}^*(s) = \epsilon\gamma/(2-2\gamma)$ for $s \in \mathcal{S}_U$.

- We let $\mathcal{M}_4$ be the set of instances that differ from $M_3$ by one entry. Let there be some state-action pair $(\bar{s}, \bar{a}) \in \mathcal{S}_U \times \mathcal{A}_U$ that transitions to one of the first $1/\epsilon$ good state with probability $\epsilon$ and a reward of 0. For simplicity, we assume that $1/\epsilon$ is an integer. Let all other transition probabilities be identical to $M_3$. There are $|\mathcal{S}_U \times \mathcal{A}_U| \times 1/\epsilon$ such instances. For $M_4 \in \mathcal{M}_4$, we have $v_{M_4}^*(\bar{s}) = \epsilon\gamma/(1-\gamma)$ and $v_{M_4}^*(s) = \epsilon\gamma/(2-2\gamma)$ for all $s \in \mathcal{S}_U/\{\bar{s}\}$.

We see that $\|v_{M_3}^* - v_{M_4}^*\|_\infty = \frac{\gamma}{2(1-\gamma)}$. For any algorithm that outputs an $\epsilon\gamma/(4-4\gamma)$-optimal policy, it should be able to differentiate $\mathcal{M}_4$ from $\mathcal{M}_3$. In the case of `CDP MDP`, differentiating $\mathcal{M}_4$ from $\mathcal{M}_3$ requires one to search for one single discrepancy in an array of size $|\mathcal{S}_U \times \mathcal{A}_U \times \frac{1}{\epsilon}| = \Omega(|\mathcal{S}||\mathcal{A}|/\epsilon)$; see Figure 2. In the case of `BinaryTreeMDP`, differentiating $\mathcal{M}_4$ from $\mathcal{M}_3$ requires one to search for two discrepancies from $\Omega(|\mathcal{S}||\mathcal{A}|/\epsilon)$ leaves; see Figure 3.

|   |   | $s_{B,1}$ | $s_{G,1}$ | $s_{B,2}$ | $s_{G,2}$ | $\cdots$ | $s_{B,1/\epsilon}$ | $s_{G,1/\epsilon}$ | $\cdots$ | $s_{B,k}$ | $s_{G,k}$ | $s_N$ |
|---|---|---|---|---|---|---|---|---|---|---|---|---|
| $M_3$ | $\mathcal{S}_U \times \mathcal{A}_U$ | $\epsilon$ | $\epsilon$ | $2\epsilon$ | $2\epsilon$ | $\cdots$ | 1 | 1 | $\cdots$ | 1 | 1 | 1 |
|   |   | $\epsilon$ | $\epsilon$ | **$2\epsilon$** | $2\epsilon$ | $\cdots$ | 1 | 1 | $\cdots$ | 1 | 1 | 1 |
|   |   | $\vdots$ | $\vdots$ | $\vdots$ | $\vdots$ | $\ddots$ | $\vdots$ | $\vdots$ | $\ddots$ | $\vdots$ | $\vdots$ | $\vdots$ |
|   |   | $\epsilon$ | $\epsilon$ | $2\epsilon$ | $2\epsilon$ | $\cdots$ | 1 | 1 | $\cdots$ | 1 | 1 | 1 |
|   | $\mathcal{S}_U \times a_N$ | 0 | 0 | 0 | 0 | $\cdots$ | 0 | 0 | $\cdots$ | 0 | 0 | 1 |
| $M_4$ | $\mathcal{S}_U \times \mathcal{A}_U$ | $\epsilon$ | $\epsilon$ | $2\epsilon$ | $2\epsilon$ | $\cdots$ | 1 | 1 | $\cdots$ | 1 | 1 | 1 |
|   |   | $\epsilon$ | $\epsilon$ | **$\epsilon$** | $2\epsilon$ | $\cdots$ | 1 | 1 | $\cdots$ | 1 | 1 | 1 |
|   |   | $\vdots$ | $\vdots$ | $\vdots$ | $\vdots$ | $\ddots$ | $\vdots$ | $\vdots$ | $\ddots$ | $\vdots$ | $\vdots$ | $\vdots$ |
|   |   | $\epsilon$ | $\epsilon$ | $2\epsilon$ | $2\epsilon$ | $\cdots$ | 1 | 1 | $\cdots$ | 1 | 1 | 1 |
|   | $\mathcal{S}_U \times a_N$ | 0 | 0 | 0 | 0 | $\cdots$ | 0 | 0 | $\cdots$ | 0 | 0 | 1 |

Figure 2: Input of `CDP MDP` : Arrays of Cumulative Probabilities for $M_3 \in \mathcal{M}_3$ and $M_4 \in \mathcal{M}_4$

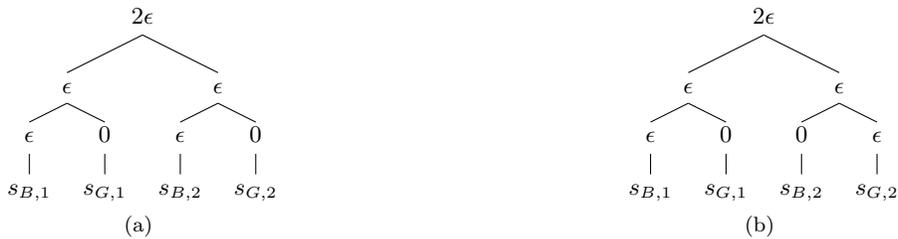

Figure 3: Input of `Binary Tree` MDP: (a) The snippet of the binary tree of $M_3 \in \mathcal{M}_3$, where all transitions are to bad states. (b) The snippet of the binary tree of $M_4 \in \mathcal{M}_4$, where there is some $(s,a)$ which transitions to a good state $s_{G,2}$ with probability $\epsilon$.

## 5 Proof of Main Theorems

In this section, we begin by introducing a sub-problem about differentiating matrices and establish its complexity lower bound. Then we develop the proofs of the main theorems by using Yao's minimax principle.

## 5.1 A Sub-Problem

To facilitate our proof, we first introduce a sub-problem and give a lemma that is vital to the proof of Theorem 1.

**Definition 4** (`Matrix Differentiation`). *Let $A$ and $B$ be two $m$ by $n$ matrices taking values of $0$ or $1$. For every row of $A$, there is an entry of $1$ and the other entries are $0$. $B$ is similar to $A$ except that there is a row with all zeros. We say that $(A, B)$ is a "pair of matrices" if $A$ and $B$ differ by exactly one entry. Note that there are $mn^m$ pairs of matrices in total. We focus on deterministic algorithms that take the $m \times n$ binary matrix as the input. For any pair $(A, B)$, we say that a deterministic algorithm is able to distinguish $A$ from $B$ if it queries the entry $(i, j)$ where $A_{ij} \neq B_{ij}$ within a given number of steps when either $A$ or $B$ is given as the input.*

**Lemma 1.** *For any deterministic algorithm that applies to `Matrix Differentiation`, it needs $\Omega(mn)$ queries to distinguish $3mn^m/5$ pairs of matrices.*

In order to prove Lemma 1, we model any deterministic algorithm for the matrix differentiation problem as a tree program (which was used in [BFK+81] to study the complexity of sorting). As the algorithm makes a new query to the input matrix, the state of the algorithm branches according to the response until the maximal number of queries is reached. The proof is given in the appendix.

## 5.2 Proof of Theorem 1

Let us summarize the main arguments of the proof: Suppose that some randomized algorithm can find an $\epsilon$-optimal policy to the MDP with high probability within a time limit. Then by using Yao's minimax principal, we obtain that there exists a deterministic algorithm that is able to successfully distinguish a large number of hard instances in $\mathcal{M}_1$ from those in $\mathcal{M}_2$. It follows that one can construct an algorithm for `Matrix Differentiation` and apply Lemma 1 to establish the lower bound. The proof has a similar spirit as that of the relational adversary method [Amb00, Aar06].

*Proof of Theorem 1.* Given the state space $\mathcal{S}$, the action space $\mathcal{A}$ and the discount factor $\gamma$, we denote $\mathcal{M}_0$ to be the set of the MDP instances where the entries of $P$ and $r$ are multiples of 0.01. Note that $\mathcal{M}_1$ and $\mathcal{M}_2$ are two subsets of $\mathcal{M}_0$. Let $\Pi_T$ be the set of all the deterministic algorithms that takes $M_0 \in \mathcal{M}_0$ as input and run at most $T$ steps (each step is an query of an entry). Let $\mu$ be a distribution on $\Pi_T$ and let $\mathcal{D}$ be a distribution on $\mathcal{M}_0$. Recall that a randomized algorithm running at most $T$ steps is a distribution on $\Pi_T$. Theorem 1 can be equivalently stated as: If

$$\max_{\mu \in \mathcal{P}(\Pi_T)} \min_{M_0 \in \mathcal{M}_0} \mathbf{P}_{\pi \sim \mu(M_0)} \left( \max_{s \in \mathcal{S}} \left| v_{M_0}^\pi(s) - v_{M_0}^*(s) \right| \leq \epsilon \right) \geq 9/10, \tag{2}$$

then $T = \Omega(|\mathcal{S}|^2|\mathcal{A}|)$ where $\mathcal{P}(\Pi_T)$ is the set of all probability measures on $\Pi_T$ and $\pi$ is the policy computed by the randomized algorithm $\mu$ on input $M_0$. If (2) holds, we apply Yao's minimax principle [Yao77] and get

$$\min_{\mathcal{D} \in \mathcal{P}(\mathcal{M}_0)} \max_{\mu \in \Pi_T} \mathbf{P}_{M_0 \sim \mathcal{D}} \left( \max_{s \in \mathcal{S}} \left| v_{M_0}^\pi(s) - v_{M_0}^*(s) \right| \leq \epsilon \right) \geq 9/10, \tag{3}$$

where $\pi$ is the policy computed by $\mu$ on input $M_0$. The advantage of using Yao's minimax principle is that we have converted the complexity of randomized algorithms into the complexity of deterministic algorithms on randomized input.

Now let $\mathcal{D}_1$ and $\mathcal{D}_2$ be the uniform distributions over $\mathcal{M}_1$ and $\mathcal{M}_2$ respectively. Let $\mathcal{D}$ be an equal mixture of $\mathcal{D}_1$ and $\mathcal{D}_2$. Assume that there exists a deterministic algorithm $\mu$ which outputs an $\epsilon$-optimal policy with probability $9/10$ when the MDP instance is drawn from $\mathcal{D}$. As a result, $\mu$ must succeed with probability at least $4/5$ if the instance is drawn from either $\mathcal{D}_1$ or $\mathcal{D}_2$ alone. Let $C_1 \subset \mathcal{M}_1$ and $C_2 \subset \mathcal{M}_2$ be the sets of instances on which $\mu$ succeeds. Let $m = |\mathcal{S}_U \times \mathcal{A}_U|$ and $n = |\mathcal{S}_B|$. By the discussion in Section 4, we have

$$|C_1| \geq \frac{4}{5}|\mathcal{M}_1| = \frac{4}{5}n^m, \quad |C_2| \geq \frac{4}{5}|\mathcal{M}_2| = \frac{4}{5}mn^m.$$

For instances $M_1 \in \mathcal{M}_1$ and $M_2 \in \mathcal{M}_2$, we say that $(M_1, M_2)$ is a *pair of MDP instances* if the following two conditions are satisfied: (i) The transition probabilities of $M_1$ and $M_2$ differ at only two entries, $(\bar{s}, \bar{a}, s')$ and $(\bar{s}, \bar{a}, \bar{s}')$. (ii) $s'$ and $\bar{s}'$ have the same index within $\mathcal{S}_B$ and $\mathcal{S}_G$ (e.g., $s' = s_{B,1}$ and $\bar{s}' = s_{G,1}$).

Consider the graph where nodes are MDP instances and there is an edge between every "pair". We can see that each $M_1 \in \mathcal{M}_1$ has degree $m$ and each $M_2 \in \mathcal{M}_2$ has degree 1. Let $e(\cdot, \cdot)$ be the indicator function where $e(M_1, M_2) = 1$ if $(M_1, M_2)$ is a pair (so there is an edge) and 0 otherwise. The total number of pairs between $C_1$ and $C_2$ is

$$\sum_{M_1 \in C_1, M_2 \in C_2} e(M_1, M_2)$$
$$\geq \sum_{M_1 \in C_1, M_2 \in \mathcal{M}_2} e(M_1, M_2) - \sum_{M_1 \in \mathcal{M}_1, M_2 \in \mathcal{M}_2 \setminus C_2} e(M_1, M_2)$$
$$\geq \frac{4}{5} n^m \times m - \frac{1}{5} m n^m = \frac{3}{5} m n^m.$$

It means that the deterministic algorithm $\mu$ must be able to distinguish at least $3mn^m/5$ pairs of MDP instances.

In what follows, we use $\mu$ to construct an algorithm that applies to `Matrix Differentiation`. We first construct an auxiliary algorithm $\mu'$ that mimics the work flow of $\mu$ with a slight difference. Suppose that $\mu$ queries an entry $(s, a, s')$. If $s'$ is a bad states, $\mu'$ will query $(s, a, s')$ and branch in the same way as if $\mu$ queries the entry $(s, a, s')$ and branches. If $s'$ is a good state, $\mu'$ will also query $(s, a, s')$, but regardless of the value of $(s, a, s')$, $\mu'$ will branch in the way that $\mu$ branches after reading a value 0 at $(s, a, s')$. The algorithm $\mu'$ outputs all the actual values of queries it read along the computation path.

We claim that $\mu'$ is able to differentiate all the pairs that $\mu$ differentiate. Let $(M_1, M_2)$ be a pair of MDP instances which can be differentiated by $\mu$. Suppose that we run both $\mu$ and $\mu'$ on $M_1$ (or $M_2$). Let $(\bar{s}, \bar{a}, \bar{s}')$ be the entry on which $M_1$ and $M_2$ differ - this entry is eventually queried by $\mu$. Let $(s, a, s')$ be any previous query made by $\mu$. If $s'$ is a good state, then the value of $(s, a, s')$ for both $M_1$ and $M_2$ must be 0 (according to the construction of $M_1, M_2$ in Section 4.1). Therefore $\mu'$ will always mimic the moves of $\mu$ until it queries the right entry to differentiate $M_1$ from $M_2$. As a result, $\mu'$ can distinguish all the pairs that $\mu$ distinguishes in $T$ steps. Now let us we construct an algorithm based on $\mu'$ that applies to `Matrix Differentiation`, which is given in Algorithm 1.

---

**Algorithm 1** The algorithm constructed from $\mu'$ that applies to `Matrix Differentiation`

1: **Input:** An $m \times n$ matrix $A$.
2: Set $|\mathcal{S}_U|, |\mathcal{S}_B|, |\mathcal{S}_G|$ and $|\mathcal{A}_U|$ such that $m = |\mathcal{S}_U \times \mathcal{A}_U|$ and $n = |\mathcal{S}_G| = |\mathcal{S}_B|$.
3: **while** $\mu'$ hasn't terminated **do**
4:     Let $(s, a, s')$ be the entry to be queried by $\mu'$
5:     Let $i$ be the index of $(s, a)$ in $\mathcal{S}_U \times \mathcal{A}_U$
6:     **if** $s' \in \mathcal{S}_B$ **then**
7:         Query $A_{ij}$ where $j$ is the index of $s'$ in $\mathcal{S}_B$
8:     **else if** $s' \in \mathcal{S}_G$ **then**
9:         Query $A_{ij}$ where $j$ is the index of $s'$ in $\mathcal{S}_G$
10:     **end if**
11:     Return $A_{ij}$ to answer the query of $\mu'$
12: **end while**

---

We claim that Algorithm 1 can distinguish at least $3mn^m/5$ pairs of matrices for `Matrix Differentiation`. Given a pair $(M_1, M_2) \in \mathcal{M}_1 \times \mathcal{M}_2$, we denote the subarray of transition probabilities corresponding to $(\mathcal{S}_U \times \mathcal{A}) \times \mathcal{S}_B$ as the matrix $A$ and the matrix $B$. Note that $A$ differs from $B$ at a single entry, so $(A, B)$ is a pair of `Matrix Differentiation`. Due to the definition for pairs of MDP instances (condition (ii)), we can verify that there is a one-to-one correspondence between the set of pairs of matrices for `Matrix`

Differentiation and the set of pairs of MDP instances. For each pair $(M_1, M_2)$ that $\mu'$ distinguishes, $\mu'$ queries an entry $(s, a, s')$ on which $M_1$ and $M_2$ differ. Because of condition(ii), Algorithm 1 will query the corresponding entry $(i, j)$ on which $A$ and $B$ differ. Since $\mu'$ differentiate $3mn^m/5$ pairs of MDP instances in running time $T$, Algorithm 1 must be able to differentiate $3mn^m/5$ pairs of matrices for Matrix Differentiation using $T$ queries. Finally, we apply Lemma 1 and obtain $T = \Omega(|\mathcal{S}|^2|\mathcal{A}|)$. ∎

### 5.3 Proofs of Theorem 2 and 3

The proofs of Theorems 2 and 3 are significantly simpler than that of Theorem 1. This is because $\mathcal{M}_3$ and $\mathcal{M}_4$ differs in one single entry. The following proofs do not rely on Lemma 1.

*Proof of Theorem 2.* Using similar notations as in Section 5.2, we restate Theorem 2 as: If

$$\max_{\mu \in \mathcal{D}(\Pi_T)} \min_{M_0 \in \mathcal{M}_0} \mathbf{P}_{\pi \sim \mu(M_0)} \left( \max_{s \in \mathcal{S}} \left| v^\pi_{M_0}(s) - v^*_{M_0}(s) \right| \leq \epsilon \right) \geq 9/10,$$

then $T = \Omega(|\mathcal{S}||\mathcal{A}|/\epsilon)$, where $\Pi_T$ is the set of all the deterministic algorithms for CDP MDP . We apply the minimax principal similarly as in Section 5.2.

Let $\mathcal{D}_3, \mathcal{D}_4$ be the uniform distributions over $\mathcal{M}_3, \mathcal{M}_4$ respectively. Let $\mathcal{D}$ be an equal mixture of $\mathcal{D}_3$ and $\mathcal{D}_4$. We say that $M_3 \in \mathcal{M}_3$ and $M_4 \in \mathcal{M}_4$ is a pair if $M_3$ and $M_4$ differs by exactly one entry. Note that $|\mathcal{M}_3| = 1$ and there are $|\mathcal{S}||\mathcal{A}|/\epsilon$ pairs in total.

Suppose that there is a deterministic algorithm $\mu$ that finds an $\frac{\epsilon\gamma}{4-4\gamma}$-optimal policy with probability at least $9/10$, it must succeed (finds the optimal action) on $M_3$ and on $4/5$ of all instances in $\mathcal{M}_4$. It means that the algorithm $\mu$ must be able to distinguish at least $4/5 \times |\mathcal{S}||\mathcal{A}|/\epsilon$ pairs. When $\mu$ queries a new entry, it distinguishes one more pair of $(M_3, M_4)$ that happen to only differ at this entry. The goal is to to distinguish at least $4/5 \cdot |\mathcal{S}||\mathcal{A}|/\epsilon$ pairs. Therefore, $\mu$ makes at least $\Omega(|\mathcal{S}||\mathcal{A}|/\epsilon)$ queries and its running time is lower bounded by $\Omega(|\mathcal{S}||\mathcal{A}|/\epsilon)$. ∎

*Proof of Theorem 3.* The proof of Theorem 3 is very similar to the proof of Theorem 2. The intuition is that for any pair $(M_3, M_4)$, the binary tree representations of $M_3$ and $M_4$ are exactly the same except for two leaves. In order to find these two leaves or to be certain that such leaves do not exist, we need to query at least $\Omega(|\mathcal{S}||\mathcal{A}|/\epsilon)$ leaves. We omit the details for brevity. ∎

# A Proof of Lemma 1

**Computational Model:** We model the deterministic algorithm solving `Matrix Differentiation` by a *tree program*. A tree program of length $n$ is a directed tree with $n + 1$ layers, where each layer represents a time step. Each node of the tree represents a state of the algorithm that leads to a query of one entry of the input, whose edges correspond to the possible responses to the query. Given an input, the query at the root is first tested, and the state of the algorithm moves following the edge given by the response to the query. This leads to the second query, the position of which is decided by the deterministic algroithm and the response of the first query. The query continues until it reaches the leaf of the tree.

For `Matrix Differentiation`, the entries of the inputs only take value of 0 or 1. Thus, every node has exactly two outgoing edges to the next layer. We assume that the edge to the left child corresponds to the response of 0 and the edge to the right child corresponds to the response of 1. Each inner node in the tree is associated with a tuple $(i, j)$, which is the entry to be queried at that node.

Before giving the proof, we give a few more definitions. For any input, the corresponding *computation path* is the sequence of nodes visited by the algorithm until termination. We define the *$i$-th slice* of the tree program to be the set of all the nodes to which the path has $i - 1$ edges labelled by 1. Recall that input matrices $A$ and $B$ are a pair if and only if $A$ and $B$ differs by exactly one entry. We say that *a node distinguishes the pair $(A, B)$* if the node is on the computation paths of both $A$ and $B$ and $A_{ij} \neq B_{ij}$ where $(i, j)$ is the entry to be queried at that node.

We are ready to introduce the auxillary lemmas that establish the properties of the tree programs for any deterministic algorithm solving the `Matrix Differentiation`. Without loss of generality, we assume that the algorithm won't query the same entry two times in a single computation path. We have the following lemmas.

**Lemma 2.** *Any single node at the $k$-th slice can distinguish at most $n^{m-k}$ pairs.*

*Proof.* Suppose that a node at the $k$-th slice distinguishes a pair $(A, B)$. It means that the node is on the computation paths of both $A$ and $B$, and the responses to the previous queries on the path are the same given either $A$ or $B$ as the input. Observe that the responses of $k - 1$ of previous queries are 1, which means that the algorithm knows at least $k - 1$ fixed rows of $A$ and $B$. Now if $A_{ij} \neq B_{ij}$ where $(i, j)$ is the query of the current node, the $i$-th row of $A$ and $B$ is also fixed and the positions of unknown ones are only in the remaining $m - k$ rows. The total number of possible configurations of these rows is at most $n^{m-k}$. As a result, any single node at the $k$-th slice can distinguish at most $n^{m-k}$ pairs. ∎

**Lemma 3.** *The nodes at the $k$-th slice can distinguish at most $n^m$ pairs in total.*

*Proof.* Let $(A, B)$ be any pair of the input where all the rows of $A$ have a 1. We want to count the number of pairs involving $A$ that is distinguished at the $k$-th slice. We claim that if the computation path of $A$ doesn't reach the $(k + 1)$-th slice, then none of the pairs involving $A$ will be distinguished at the $k$-th slice. Suppose that the computation path doesn't reach the $k$-th slice. It is obvious that none of such pairs will be distinguished at $k$-th slice. If the computation path reaches the $k$-th slice and stays on it, then all the responses at the $k$-th slice are 0. Yet, if the tree program distinguishes a pair $(A, B)$ at the $k$-th slice, it must query an entry $(i, j)$ where $A_{ij} = 1$ and $B_{ij} = 0$. As a result, if there exists a pair involving $A$ that is distinguished at the $k$-th slice, the computation path of $A$ must reach the $(k + 1)$-th slice.

Assume that the computation path of $A$ reaches the $(k + 1)$-th slice. We can see that the path will reach $(k + 1)$-th slice from some node. Denote $(i, j)$ to be the entry queried at that node. Observe that the node can distinguish only one pair $(A, B)$ of all the $m$ pairs involving $A$ as there is only one possible $B$ such that $A_{ij} \neq B_{ij}$. It means that for every possible $A$, the tree program distinguish at most one pair involving $A$ at the $k$-th slice. As there are at most $n^m$ configurations of $A$, the tree program can distinguish at most $n^m$ pairs at the $k$-th slice. ∎

The intuition of Lemma 3 is that whenever we observe one more "1" in a query, the algorithm can distinguish at most $n^m$ more pairs. Now, we give our third lemma.

**Lemma 4.** *Let $L \geq m$ be the number of layers of the tree program. Then the total number of nodes at the $k$-th slice is less than or equal to $\binom{L}{k}$.*

*Proof.* Each node at the $k$-th slice is uniquely identified by the path from the root to the node, which can be represented by a sequence of numbers $(a_1, a_2, \ldots, a_k)$. Here, $a_i$ is the number of nodes on the path which belongs to the $i$-th slice. The number of possible paths is at most the number of integer solution to the equation $\sum_{i=1}^{k} a_i \leq L$. We introduce a slack variable $\alpha$ so that $\alpha + \sum_{i=1}^{k} a_i = L+1$ where $\alpha \geq 1, a_1 \geq 1, \ldots, a_k \geq 1$. It is easy to see that the number of solutions is $\binom{L}{k}$. ∎

The above three lemmas establish the properties of the tree program for the `Matrix Differentiation`. We are now ready to prove the main lemma.

**Proof of Lemma 1.** Assume that we have a deterministic algorithm that distinguishes at least $3mn^5/5$ pairs in $nk$ time steps, where $k$ is an integer smaller than $m/5$. We first construct the tree program of the deterministic algorithm. We divide the tree into two parts, where the first part is the first $5k$ slices and the second part is the remaining $m - 5k$ slices. By Lemma 3, the tree program can distinguish at most $5kn^m$ pairs in the first $5k$ slices. We then bound the number of pairs which the tree can distinguish in the remaining slices. In the $i$-th slice, there are at most $\binom{nk}{i}$ nodes by Lemma 4. By Lemma 2, each node can distinguish at most $n^{m-i}$ pairs. As a result, the tree program can distinguish at most $\binom{nk}{i}n^{m-i}$ pairs at $i$-th slice for $i > 5k$. The total number of pairs we can distinguish in the remaining slices is then

$$\sum_{i=5k+1}^{m} \binom{nk}{i} n^{m-i} \leq \sum_{i=5k}^{m} \binom{nk}{i} n^{m-i} \leq \sum_{i=5k}^{m} \frac{(nk)^i}{i!} n^{m-i} = n^m \sum_{i=5k}^{m} \frac{k^i}{i!}.$$

By Taylor's remainder theorem, we have

$$\sum_{i=5k}^{m} \frac{k^i}{i!} \leq \frac{f^{(5k)}(k)}{(5k)!} k^{5k} = \frac{e^k k^{5k}}{(5k)!},$$

where $f(k) = e^k$. By using Stirling's formula where $(5k)! \geq \sqrt{2\pi}(5k)^{5k+1/2}e^{-5k}$, we have the following

$$n^m \sum_{i=5k}^{m} \frac{k^i}{i!} \leq \frac{n^m e^k k^{5k}}{\sqrt{2\pi}(5k)^{5k+1/2}e^{-5k}} \leq n^m \left(\frac{e^{6/5}k}{5k}\right)^{5k} \leq n^m.$$

As a result, the total number of pairs we can distinguish is smaller than $(5k+1)n^m$ when the layer of the tree program is $nk$. In order to distinguish at least $3mn^5/5$ pairs, the number of layers of the tree program should be $\Omega(mn)$. ∎